\documentclass[sigconf,natbib=true,anonymous=false]{acmart}
\usepackage{multirow}

\AtBeginDocument{%
  \providecommand\BibTeX{{%
    \normalfont B\kern-0.5em{\scshape i\kern-0.25em b}\kern-0.8em\TeX}}}

\begin{document}
\fancyhead{}
\title{Interactive Query Clarification and Refinement via User Simulation}

\author{Pierre Erbacher}
\affiliation{%
  \institution{ISIR - Sorbonne University}
  \city{Paris}
  \country{France}}
\email{pierre.erbacher@isir.upmc.fr}

\author{Ludovic Denoyer}
\affiliation{
  \institution{Sorbonne University}
  \city{Paris}
  \country{France} \begin{tiny} Currently working at Ubisoft \end{tiny} }
\email{ludovic.denoyer@sorbonne-université.fr}

\author{Laure Soulier}
\affiliation{
  \institution{ISIR - Sorbonne University}
  \city{Paris}
  \country{France}}
\email{laure.soulier@isir.upmc.fr}

\begin{abstract}
When users initiate search sessions, their queries are often unclear or might  lack of context; this resulting in inefficient document ranking.
Multiple approaches have been proposed by the Information Retrieval community to add context and retrieve documents aligned with users' intents. While some work focus on  query disambiguation using users' browsing history, a recent line of work proposes to interact with users by asking clarification questions or/and proposing clarification panels. However, these approaches count either a limited number (i.e., 1) of interactions with user or log-based interactions. In this paper, we propose and evaluate a fully simulated  query clarification framework allowing multi-turn interactions between IR systems and user agents.
\end{abstract}



\maketitle

\section{Introduction}
Understanding information need is a long-standing issue in Information Retrieval (IR) \cite{Cronen-Townsend-ambiguity,Jansen-ambiguity,Sanderson-ambiguity}, often highlighted by the difficulty for users to formulate open-ended information needs into queries. Queries are thus often under-specified and/or ambiguous \cite{Jansen-ambiguity}. 
Depending on the user or the context, the same query may refer to different intents. For instance, the query "Orange" may refer to several topics, including company names, locations, songs titles, ...
To tackle this issue, numerous works have been proposed. One first line of work relies on query reformulation \cite{Amati-reformulation,Lavenkro-reformulation,Rocchio-SMART-1971,zukerman-raskutti-2002-lexical} where the objective is to rewrite the query. A lot of effort has been provided by designing models based on either (pseudo-)relevance feedback \cite{Amati-reformulation,Lavenkro-reformulation,Rocchio-SMART-1971} or external knowledge resources \cite{zukerman-raskutti-2002-lexical}. But recently, the advances in machine translation models and in large language models have turned this task into a query generation task \cite{Dalton-CAST,elgohary-etal-2019-unpack}. 
Another category of work focuses on search/query diversification \cite{Agrawal-diversity,Fei2016-diversification, 10.1145/290941.291025, MacAvaney-IntentT5, Nogueira2019MultiagentQR} to increase the query coverage, particularly when the query is multi-faceted. 
While early search result diversification work have designed or derived the Maximal Margin Relevance (MMR) model \cite{10.1145/290941.291025}, other techniques such as document clustering \cite{Agrawal-diversity} or document-driven voting scheme \cite{dang-diversity} has been used to retrieve a diversified list of documents.
Recently, MacAvaney et al. \cite{MacAvaney-IntentT5} have proposed to focus on query diversification by generating queries by designing a Distributional Causal Language Modeling. However, for all these diversification techniques, the issued document list might include some top-ranked documents that do not match with the user's intent \cite{Wang-limitdiversification}. This highlights the need to clarify users' queries before retrieving documents. 
A last category of work aims to leverage search history to infer user's profile or session context, with the objective to ground the initial query  \cite{Matthijs-UserHistory, Weize-PreSearchContext,Harvey2013BuildingUP, Xiang2010ContextawareRI, Bennett2012ModelingTI}. By encoding short-term behavior or long-term behavior as features of the ranking model, these methods manage to retrieve more personalized documents. But, these approaches might be limited by the user's behavior's variability over time \cite{Analyzing-clarification}. 
In parallel,  neural ranking models, such as \cite{Khattab-Colbert,Hofstatter-topicaware}, have  increased their performance due to the fine-grained capture of queries and documents semantics. Nevertheless, the query ambiguity is often inherent to users, increasing the need to place the user at the center of the IR process.

A promising approach has been proposed in \cite{Aliannejadi19-askingclarifying} to clarify information needs by proactively interacting with the user. The authors propose a conversation framework that consists in generating clarifying questions when the query is ambiguous. Clarifying questions might be query reformulation (e.g., "Would you like to know how to care for your dog during heat?" for the initial query "dog heat" as in \cite{Aliannejadi19-askingclarifying}) or questions with possible options (e.g., "what do you want to know about this british mathematician? Options: movie, suicide note, quotes, biography" for the initial query "alan turing" as in \cite{Zamani-Clarification}). With this in mind, the classic workflow for asking clarifications is based on three main steps \cite{Aliannejadi19-askingclarifying}:  1) the IR system produces a clarifying question for the user, 2) the latter provides an answer or selects an option, and 3) the IR system ranks documents according to the user's feedback. The pioneering work \cite{Aliannejadi19-askingclarifying} aims at generating clarifying questions by 1) retrieving a predefined set of questions using a Bert-based model and 2)  at each turn, selecting the best query through a conversation history-driven model. The user's answer corresponds to a predefined text built using crowd-sourcing.   One drawback of this approach is that the multi-turn conversation is log-based, interactively simulated using predefined logs of conversation history (i.e., sequence of questions/answers obtained by HITS). This simulated conversation defined a priori without interaction with the proposed question selection model might hinder the evaluation performance in the sense that we are not sure about the soundness of the conversation flow.
Other work \cite{Zamani-Clarification,Wu2018QuerySW,Guo2011IntentawareQS,Santos2012LearningTR,10.1145/2348283.2348290} tackle this issue by proposing generative models, that create clarification question or query suggestions.  But they do not address the multi-turn framework, stopping the clarification process at the first interaction.

In this paper, we propose to build a fully simulated query clarification framework allowing multi-turn interactions between IR and user agents. Following \cite{Aliannejadi19-askingclarifying}, the IR agent identifies candidate queries and ranks them in the context of the user-system interactions to clarify the initial query issued by the user agent. We particularly target simple information needs, multiple information needs are let for future work since it might impact the modeling of the query ranking function. Our framework can be seen as a basis and a proof-of-concept for future work willing to integrate sequential models (namely reinforcement learning models) for question clarification. It is worth noting that large language models relying on attention mechanisms (transformers) are not yet well suited to handle sequential interactions and long-term planning, as current models are hardly trainable with current reinforcement learning algorithms \cite{Chen2020ExploringFQ}. Therefore, all agent components in our framework are based on continuous and simple models. 
To validate our simulation framework, we conduct an experimental analysis on the MS Marco dataset. We show the benefit of multi-turn interactivity and evaluate the effectiveness of different question selection strategies.

\section{Question Clarification Simulation Framework}

\subsection{Overview and Research Hypotheses}
\begin{figure}[t]
    \centering
    \includegraphics[width=0.45\textwidth]{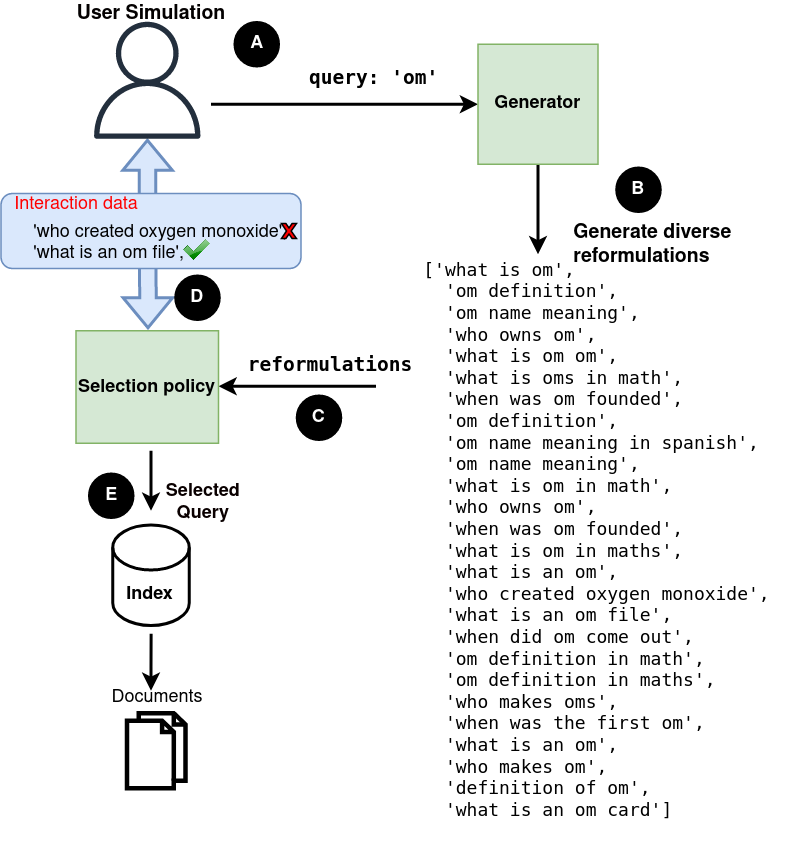}
    \vspace{-0.6cm}
    \caption{Query clarification simulation framework}
    \vspace{-0.4cm}
    \label{fig:pipline}
\end{figure}

Our query clarification simulation framework is inspired from \cite{Aliannejadi19-askingclarifying}, but provides the possibility of leveraging user-system agents' interactions sequentially.
More particularly, our  framework is illustrated in Figure \ref{fig:pipline} and relies on the following workflow: 
\begin{itemize}
    \item A) The user issues an initial query $q_0$ associated to her/his information need $i$ to the IR system.
    \item B) The IR system generates a set $Q=\{q_1,q_2,...,q_m\}$ of candidate queries which might express different query reformulations or diversified queries to better explore the information need $i$. 
    \item C) The IR system selects $N$ queries to display to the user. To do so, we propose to follow \cite{Aliannejadi19-askingclarifying} and design a model    ranking the candidate query set $Q$ to identify the top $N$ queries.
    \item D) The user selects one of the $N$ queries, enabling to extract positive and negative feedback, resp. noted $(q^+,q^-)$. 
    \item Steps C) and D) can be repeated several times to model  multi-turn interactions. The query set ranking function (step C) integrates the user's sequential feedback ($q^+,q^-$) to improve the query ranking along with the interaction simulation. 
    \item E) After $T$ turns, the IR system considers the best ranked query as the optimal query reformulation and runs a ranking model to retrieve documents.
\end{itemize}

The design of this evaluation framework is guided by some choices/research hypothesis. \\
$\bullet$ First, following \cite{Aliannejadi19-askingclarifying}, we consider a fixed set of candidate queries $Q=\{q_1,q_2,...,q_m\}$ constituting the reformulation of the initial query  $q_0$. All the interactions are leveraged to improve step by step the ranking of this candidate query set so as, at the end of the session, the final query used for retrieving documents is a good clarification of its initial one. Obviously, this means that the set of candidate queries includes a large variety of queries which, for some of them, improve the search performance. \\
$\bullet$ Second, following \cite{Zamani-Clarification}, we propose to model question clarification as a possible option between two reformulated queries. In other words, expressed in natural language, the IR system agent would ask the user agent the following question: "Which reformulated query do you prefer? A or B". This implies that the user is willing to judge queries A or B regarding its information need.\\
$\bullet$ Third, guided by the motivation to propose a framework for future work on sequential models, we consider here that each agent component is modeled at the embedding level. Indeed, leveraging large language models for generating/ranking questions is very effective, but integrating them into reinforcement learning models is still challenging (one main reason being the computational cost). This means that we processed \textit{a priori} all queries and documents to represent them using text embeddings. This processing is done offline, alleviating the sequential modeling of the text encoding.

In what follows, we present the different components behind the IR system and user agents.

\subsection{The IR  System Agent}
The IR  agent has three objectives in our framework: 1) generating the set of candidate reformulated queries willing to be presented to the user, 2) ranking this set to identify the most relevant queries according to the interaction history, 3) ranking documents using the best-ranked query (ending the interactive session). 

\paragraph{Generation of the  candidate reformulated query set.}
The objective here is to instantiate various and diverse reformulation covering a wide range of relevant topics for the initial query $q_0$. Different techniques might be used, leveraging large language models \cite{Nogueira2019DocumentEB,Raffel-T2T,RaoD19-adversarial}, query diversification \cite{Fei2016-diversification,MacAvaney-IntentT5,ye-etal-2021-one2set} or query expansion \cite{Pal-QE}. 
We propose here to use the T5 model \cite{Raffel-T2T} which is designed to translate token sequences into other token sequences. It has already been used for query reformulation tasks, demonstrating its ability for our approach \cite{Chen2020ExploringFQ,Raffel-T2T,Lin2020TREC2N}. On the top of that model, the generation process is driven by  beam diversity \cite{VijayakumarCSSL16-DiverseBeamSearch} which aims at generating a set $Q$ of diversified query reformulation, $Q=\{q_1,q_2,...,q_m\}$.

\paragraph{Ranking of queries based on the interaction history.}
The role of the selection policy is to select queries used to interact with the user agent. Following \cite{Aliannejadi19-askingclarifying} which proposes to rank queries according to both performance criteria and the interaction context, we use a conditional ranker \cite{BurgesL2R} which computes a pairwise score between two candidates queries given the context, namely the initial query $q_0$ and the additional information provided by interaction with the user. 
Let $q_i$ and $q_j$ be the candidate queries with their supervised effectiveness scores, resp. $y_i$, $y_j$.  
 The ranking model relies on:
\begin{equation}
P(y_i > y_j|q, q_i, q_j, feedback_{t-1},...,feedback_{1})
\end{equation}
For sake of simplicity, we assume that each query (initial or candidate) are represented through text embeddings. In the following, $q$ refers to query embedding and $d$ to document embedding.

In practice, the ranking model estimates a score for each query $q_i$ and $q_j$ given all the context, $\{q, q_i, q_j, feedback_{t-1},..., feedback_{1}\}$ and then compare these scores to identify which one is the most relevant.
$feedback_{t}$ corresponds to selected or not selected queries (resp. $q^+$ and $q^-$) by the user agent at interaction turn $t$. These queries are concatenated as follows: $feedback_{t}=(q^+,q^-)$ and feedback overall interaction turns are aggregated, the whole process using a Hierarchical Recurrent neural network (RNN) to encode at the interaction level and also the sequence of interactions.
Note that queries $q^+$ and $q^-$ are encoded differently using resp. a $cosin$ and $sin$ function. Moreover, we do not encode the position in which each query is presented to the user agent, as this latter does not have position bias on the clarification query selection.

\paragraph{Final ranking of documents}
Documents are retrieved with the top ranked query using a Dense Retriever model \cite{Hofstatter-topicaware}.

\subsection{The User Agent}
After issuing the initial query $q_0$, the user agent interacts with the IR system agent to refine her/his information need.
With this in mind, we hypothesize that the user is greedy toward her/his intent and fully cooperative. Therefore, he always selects the preferred query as the most similar to the intent. Despite  being unrealistic, we ignore the click bias problem for clarification panel presented in \cite{Zamani-userInteractionsForClarification, Analyzing-clarification} (as mentioned earlier). Other choices for user simulation could be done, following \cite{DBLP:journals/corr/abs-2201-11181}, but we let these variations for future work.

In practice, let $d$ be a user intent, $q_i$ and $q_j$ the clarification queries presented to the user agent. The user agent selects the best query (noted $q^+$ for highlighting positive feedback from the user) according to a similarity metric (in our case, the dot product) between the proposed queries $q_i$ and $q_j$ and  intent $d$:
\begin{equation}
q^+ = argmax_{q_i} ( \langle q_i , d \rangle  )
\end{equation}
The non-selected query, $q^-$,  expresses negative feedback.

\section{Evaluation Protocol}
Evaluating our simulation framework  consists in measuring the effectiveness of the final ranking after $T$ clarification interactions. Since the user behavior is greedy and follows a simple behavior dependent on the query selection process, the effectiveness results mainly denote the quality of this latter component. Other components (candidate set generation and final document ranking) do not depend on the interaction feedback, so we mainly focus on understanding whether the selection policy integrates users' feedback and takes good decisions to select the $N$ clarification questions. \\
For reproducibility, the code of our simulation framework and the evaluated baselines/scenarios will be released upon acceptance.

\subsection{Dataset}
We carry out our experiments on MS Marco 2020 passages which regroups 8.8M passages and more than 500K Query-Passage relevance pairs. 
Following \cite{Nogueira2019MultiStageDR}, we evaluate our model on 2 sets. The  small test set  (43 queries) and a subset of the dev set (1000 queries sampled from 59 000).  One motivation to consider these two datasets is their difficulty level: in the dev set, only one passage per query is labeled relevant in the ground truth, while several passages are considered as relevant in the test set.

\subsection{Baselines and Scenarios}
To evaluate the effectiveness of our selection policy component, we compare with: \\
\indent (1) Non-interactive settings to show the gain of interacting with users. We measure the ranking effectiveness of the user's initial user query (noted \textbf{User Query}), the \textbf{Best Reformulation} in the candidate query set - which can be seen as an oracle, and the \textbf{MonoT5} Documents re-ranker which acts as a strong ranking baseline \cite{Pradeep-monoduo}. This model is a pointwise ranking, estimating relevance scores for query/document pairs. This model  relies on a large sequence to sequence language model pretrained on various task \cite{Raffel-T2T}. Please note that using this model for the selection policy, and therefore integrating user's feedback, is not obvious since this is a seq2seq pointwise model, labels associated with queries are binary (relevant or not) and has to be grounded relative to a value. For that reason, we only consider its non-interactive scenario.\\
\indent (2) \textbf{Naive interactive selection}: At each step, we select the 2 top ranked queries from the current query rank and then remove the query which has not been selected by the user agent. The re-ranking of the candidate query set is only carried out once, at the beginning of the session, and  the size of this list decreases with the interaction number. 

\noindent To instantiate the selection policy after each interaction-driven query ranking step (step C in Figure \ref{fig:pipline}), we consider these scenarios:\\
\indent (1) \textbf{Interact. + Random Sampl}: we sample 2 queries from the ranked candidate query set to constitute the interaction pair.
    
\indent (2) \textbf{Interact. + Top 2}: we select the top 2 query reformulations at each turn.
    
\indent (3) \textbf{Interact. + random sampl@5}: we randomly select 2 queries among the  top 5 query reformulations at each turn.
    
\indent (4) \textbf{Interact. + Kmeans selection}: At each turn, queries in the candidate set are clustered in 2 groups using Kmeans. Queries from each cluster are ranked by the model. The best-ranked query from each cluster is then selected for interaction with the user. The cluster of the query not selected by the user is removed for the next turn from the set of candidate queries. This strategy corresponds to a refinement strategy, removing a group of semantically similar queries that have not been chosen by the user and going deeper in the other cluster.

\begin{figure}[t]
\centering
  \includegraphics[width=1\linewidth]{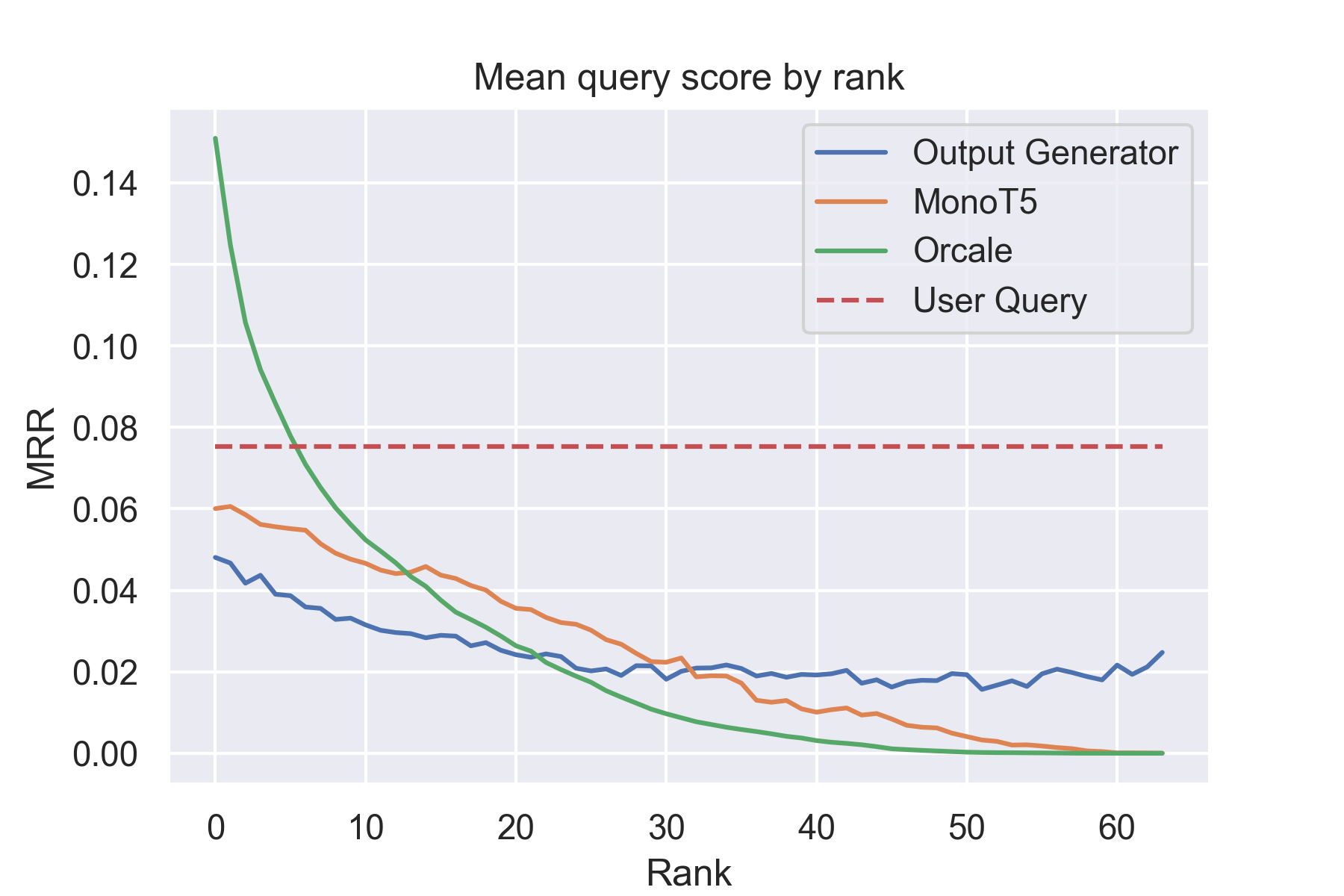}
  \vspace{-0.6cm}
  \captionof{figure}{Effectiveness score of query reformulation by rank.}
  \vspace{-0.4cm}
  \label{fig:test2}
\end{figure}

\subsection{Model Implementation}
All queries and passages embeddings are pre-computed using the Dense Retriever  proposed by \cite{Hofstatter-topicaware}. Embeddings are stored and indexed using faiss HSWN32 index \cite{faiss}.
The candidate query set is generated by diversity beam with a group penalty equal to $0.6$. The size of the candidate set is of 64. 
The number of queries displayed to the user agent is  set to $N = 2$. 
For the model hyper-parameters, we use batches of size 128, the optimizer is Adam ($\beta_1 = 0.9,\beta_2 = 0.99$) with weight decay $( = 0.01)$. We use batch normalization and dropout $(p=0.3)$ between each layer. The learning rate is set to $1 \times 10^{-4}$.

\section{Results}
\subsection{Preliminary Analysis}

We present here a preliminary analysis to quantify the potential retrieval performance gain of the candidate query set within the question clarification step. 
To do so, we compare the performance of different query rankings: 1) the candidate query generated by the T5 model ranked by decreasing order of likelihood  resulting from the Diversity Beam search  (without application of our ranking function);  2) the Oracle corresponding to the candidate query set ranked in a decreasing order according to their performance according to Mean Marginal Rank metric in the ground truth. We emphasized that this Oracle curve shows the performance of our T5 model at generating search oriented reformulations. 3) The MonoT5 ranking corresponding to candidate queries re-ranked by MonoT5.
Figure \ref{fig:test2} illustrates the performance of queries depending on their rank in the different mentioned lists. 
From Figure \ref{fig:test2}, we can see that ranking queries with MonoT5 allows to  improve the performance for the top $k$ queries (MonoT5 vs. Output Generator). This has a negative effect for the end of the list, but it is not critical in our case, since we consider selection policy regarding the top query list. Moreover, one can notice that, although performance are increased,  there is still a gap between the curve of the MonoT5 ranked list and the Oracle curve. Our intuition is that leveraging users' interactions will lower this gap, which leads to the evaluation we performed in what follows.

\begin{table}[t]
 \centering
 \resizebox{1.05\linewidth}{!}{
\begin{tabular}{lccccccc|}
\cline{3-8}
 & \multicolumn{1}{l|}{} & \multicolumn{1}{l|}{No interaction} & \multicolumn{1}{l|}{1} & \multicolumn{1}{l|}{2} & \multicolumn{1}{l|}{3} & \multicolumn{1}{l|}{4} & \multicolumn{1}{l|}{5} \\ \hline
\multicolumn{1}{|l|}{\multirow{2}{*}{User Query}} & mrr@10 & 0.4554 & - & - & - & - & - \\
\multicolumn{1}{|l|}{} & map@10 & 0.3382 & - & - & - & - & - \\ \hline
\multicolumn{1}{|l|}{\multirow{2}{*}{Best Reformulation}} & mrr@10 & 0.8720 & - & - & - & - & - \\
\multicolumn{1}{|l|}{} & map@10 & 0.5646 & - & - & - & - & - \\ \hline
\multicolumn{1}{|l|}{\multirow{2}{*}{Monot5 (query ranker)}} & mrr@10 & 0.4713 & - & - & - & - & - \\
\multicolumn{1}{|l|}{} & map@10 & 0.3209 & - & - & - & - & - \\ \hline
\multicolumn{1}{|l|}{\multirow{2}{*}{Naive selection}} & mrr@10 & 0.2135 & 0.3270 & 0.3597 & 0.4036 & 0.4191 &  0.4271 \\
\multicolumn{1}{|l|}{} & map@10 & 0.1222 & 0.1943 & 0.2205 & 0.2553 & 0.2688 & 0.2766 \\ \hline
\multicolumn{1}{|l|}{\multirow{2}{*}{Interact. + random sampl}} & mrr@10 & 0.4031 & 0.4786 & 0.4814 & 0.4903 & 0.4814 & 0.5019 \\
\multicolumn{1}{|l|}{} & map@10 & 0.2531 & 0.3344 & 0.3413 & 0.3529 & 0.3480 & 0.3685 \\ \hline

\multicolumn{1}{|l|}{\multirow{2}{*}{Interact. + Top 2
}} & mrr@10 & 0.4031 & 0.4746 & 0.4693 &  0.4903 & 0.4786 &  0.5019 \\
\multicolumn{1}{|l|}{} & map@10 & 0.2531 & 0.3294 & 0.3436 & 0.3520 & 0.3471 & 0.3428 \\ \hline

\multicolumn{1}{|l|}{\multirow{2}{*}{Interact. + random sampl@5
}} & mrr@10 & 0.4031 & 0.4734 & 0.4670 &0.4903 & 0.4798 & 0.5019\\
\multicolumn{1}{|l|}{} & map@10 & 0.2531 & 0.3287 & 0.3420 & 0.3517 & 0.3469 & 0.3451 \\ \hline
\multicolumn{1}{|l|}{\multirow{2}{*}{Interact.  + Kmean}} & mrr@10 & 0.4031 & 0.5232 & 0.4658 & 0.4692 & 0.4863 & 0.5515 \\
\multicolumn{1}{|l|}{} & map@10 & 0.2531 & 0.3706 & 0.3207 & 0.3402 & 0.3181 & 0.3347 \\ \hline
\end{tabular}
}
\caption{Effectiveness results on the Test set of MS Marco passage 2020 (43 queries - multiple relevant documents per query)}
\vspace{-0.8cm}
\label{table1}
\end{table}

\begin{table}[t]
 \centering
 \resizebox{1.05\linewidth}{!}{
\begin{tabular}{lccccccc|}
\cline{3-8}
 & \multicolumn{1}{l|}{} & \multicolumn{1}{l|}{No interaction} & \multicolumn{1}{l|}{1} & \multicolumn{1}{l|}{2} & \multicolumn{1}{l|}{3} & \multicolumn{1}{l|}{4} & \multicolumn{1}{l|}{5} \\ \hline
\multicolumn{1}{|l|}{\multirow{1}{*}{User Query}} & mrr@10 & 0.2090 & - & - & - & - & - \\ \hline
\multicolumn{1}{|l|}{\multirow{1}{*}{Best Reformulation}} & mrr@10 & 0.4119 & - & - & - & - & - \\
\hline
\multicolumn{1}{|l|}{\multirow{1}{*}{Monot5 (query ranker)}} & mrr@10 & 0.1557 & - & - & - & - & - \\
\hline
\multicolumn{1}{|l|}{\multirow{1}{*}{Naive selection}} & mrr@10 & 0.1228 & 0.1513 & 0.1659 & 0.1767 & 0.1866 &  0.1911 \\ \hline

\multicolumn{1}{|l|}{\multirow{1}{*}{Interact. + random sampl}} & mrr@10 & 0.1719 & 0.2012 & 0.1990 & 0.1954 & 0.2003 & 0.2016 \\\hline
\multicolumn{1}{|l|}{\multirow{1}{*}{Interact. +  Top 2}} & mrr@10 & 0.1719 & 0.2020 & 0.1987 & 0.1973 & 0.2017 & 0.1990 \\\hline
\multicolumn{1}{|l|}{\multirow{1}{*}{Interact. + random sampl@5}} & mrr@10 & 0.1719  & 0.2020 &  0.1983  &  0.1966 & 0.2007 &  0.2008 \\\hline
\multicolumn{1}{|l|}{\multirow{1}{*}{Interact.  + Kmean }} & mrr@10 & 0.1719 & 0.1748 & 0.1984 & 0.2016 & 0.2158 & 0.2224 \\\hline
\end{tabular}
}
\caption{Effectiveness results on the subset of MS Marco passage 2020 dev set (1000 queries - 1 relevant document per query)}
\vspace{-1cm}
\label{table2}
\end{table}

\subsection{Effectiveness Results}
We analyze here the performance of the query ranker at different interaction turns using mmr@10 and map@10. Tables \ref{table1} and \ref{table2} resp. show the results on the MS Marco passage 2020 test set and  dev set. From a general point of view, we can see that performance metrics are lower for the dev set than for the test set. This can be explained by the task difficulty, which is higher for the dev set in which only one document per query is assessed as relevant. By comparing all baselines and scenarios, we can outline the following trends. 1) The first candidate query ranking within our interactive models (No interaction columns) provides lower performance than non-interactive baselines. For instance, the \textbf{Interact. + Top2} scenario observes a decrease of 12\% in terms of mrr@10 for the test set w.r.t. the initial user query.
2) But this trend is reversed with each interaction turn to obtain for certain scenarios performance higher than baseline ones (see all interaction models in the test set, and the \textbf{Interact + Kmeans} for the dev set). 3) The interaction model with Kmean strategy looks to be the best selection policy for question clarification since it obtains the highest mrr@10 for both datasets. This is somehow intuitive because this strategy might correspond to a refinement strategy, going deeper and deeper into clusters. This is also connected with the dataset peculiarity since MS Marco is mainly composed mono-faceted questions in natural language. 

\section{Conclusion and perspectives}
This exploratory work focuses on sequential click-based interaction with a user simulation for clarifying queries. We provide a simple and easily reproducible framework simulating multi-turn interactions between a user and a IR system agent. The advantage of our framework is the simplicity of interaction, as there is no need for dataset of real and annotated user-system interactions. Experiments highlight performance gain in terms of document retrieval through the multi-turn query clarification process and provide a comparative analysis of selection strategies.  The next steps for this work are: 1) leveraging reinforcement learning for the selection policy. 2) test more diverse and more sophisticated user simulation, as done in \cite{DBLP:journals/corr/abs-2201-11181} for multi-faceted information needs.

\section{Acknowledgments}
We would like to thank the ANR JCJC project SESAMS (Projet-ANR-18-CE23-0001) for supporting Pierre Erbacher and Laure Soulier from Sorbonne Univeristé in this work.

\bibliography{bib}
\bibliographystyle{ACM-Reference-Format}
\end{document}